\documentclass[%
twocolumn,
superscriptaddress,
amsmath,amssymb,
aps,
prx,
]{revtex4-2}

\usepackage[utf8]{inputenc}
\usepackage[english]{babel}
\usepackage[T1]{fontenc}
\usepackage{amsmath}
\usepackage{hyperref}
\hypersetup{
    colorlinks = true}
\usepackage{lipsum}
\usepackage{amsfonts}

\usepackage{tikz}
\usepackage{lipsum}

\begin{document}

\title{Unsupervised learning of anomalous diffusion data}

\author{Gorka Mu\~noz-Gil}
\email{munoz.gil.gorka@gmail.com}
\affiliation{ICFO -- Institut de Ci\`encies Fot\`oniques, The Barcelona Institute of Science and Technology, Av. Carl Friedrich Gauss 3, 08860 Castelldefels (Barcelona), Spain}

\author{Guillem Guigó i Corominas}%
\affiliation{ICFO -- Institut de Ci\`encies Fot\`oniques, The Barcelona Institute of Science and Technology, Av. Carl Friedrich Gauss 3, 08860 Castelldefels (Barcelona), Spain}

\author{Maciej Lewenstein}
\affiliation{ICFO -- Institut de Ci\`encies Fot\`oniques, The Barcelona Institute of Science and Technology, Av. Carl Friedrich Gauss 3, 08860 Castelldefels (Barcelona), Spain}
\affiliation{ICREA, Pg. Llu\'is Companys 23, 08010 Barcelona, Spain}

\begin{abstract}
The characterization of diffusion processes is a keystone in our understanding of a variety of physical phenomena. Many of these deviate from Brownian motion, giving rise to anomalous diffusion. Various theoretical models exists nowadays to describe such processes, but their application to experimental setups is often challenging, due to the stochastic nature of the phenomena and the difficulty to harness reliable data. The latter often consists on short and noisy trajectories, which are hard to characterize with usual statistical approaches. In recent years, we have witnessed an impressive effort to bridge theory and experiments by means of supervised machine learning techniques, with astonishing results. In this work, we explore the use of unsupervised methods in anomalous diffusion data. We show that the main diffusion characteristics can be learnt without the need of any labelling of the data. We use such method to discriminate between anomalous diffusion models and extract their physical parameters. Moreover, we explore the feasibility of finding novel types of diffusion, in this case represented by compositions of existing diffusion models. At last, we showcase the use of the method in experimental data and demonstrate its advantages for cases where supervised learning is not applicable.
\end{abstract}

\maketitle

Stochastic diffusion processes are ubiquitous in nature, with applications in diverse fields such as physics, chemistry, biology or social sciences.
The best-known model for these is Brownian motion~\cite{metzler2019brownian}, but recently, deviations from such paradigmatic phenomena have been discovered in a huge range of systems~\cite{metzler2014anomalous}.
We define as anomalous diffusion any of these deviations, which are usually characterized by a power-law scaling of the mean squared displacement (MSD),
\begin{equation}
    \langle x^2(t) \rangle \sim t^\alpha,
\end{equation}
where $\alpha$ is defined as the anomalous exponent. For $\alpha=1$ one typically recovers normal diffusion, and usually a return to the Brownian motion universality class (for exceptions, see Ref.~\cite{wang2009anomalous}). Any $\alpha\neq1$ showcases anomalous diffusion, while other features such as non-ergodicity or correlated displacements may also be signals of its appearance~\cite{meroz2015toolbox, kindermann2017nonergodic}.

In recent years, there has been an increasing interest in the topic, motivated by the development of single particle tracking techniques (SPT), which open the possibility of studying the motion of particles beyond the diffraction limit~\cite{manzo2015review}. Such achievements have shown that diffusion in many different biological scenarios is indeed anomalous~\cite{hofling2013anomalous}. Moreover, researchers have satisfactorily applied such theoretical framework to systems at all scales, ranging from ultra cold atom experiments~\cite{kindermann2017nonergodic}, to animals~\cite{paiva2021scale} or economic signals~\cite{plerou2000economic}.

While currently both theory and experimental techniques are very well developed, connecting these two is a non-trivial task. Due to the stochastic nature of the problem, the analysis of anomalous diffusion data and its comparison to theoretical predictions often relies on the use of ensemble averages~\cite{burnecki2015estimating, kepten2015guidelines} or extensive statistical analysis~\cite{krapf2019spectral, thapa2020leveraging}. Moreover, trajectories arising from SPT are usually short and cannot be compared to the long time predictions theories usually make. On top of that, the experimental data is often very noisy and rather difficult to gather, hence increasing even more the difficulty of the problem.  

Lately, considerable interest has been devoted to develop and improve single trajectory methods, i.e. tools which extract the maximum amount of information from a single anomalous diffusion trajectory. Here, machine learning (ML) approaches have shown incredible success and are able to beat state of the art methods in a variety of scenarios~\cite{munoz2021objective}. A wide range of ML architectures have been tested: from usual neural networks~\cite{gentili2021characterization}, convolutional ~\cite{granik2019single} and recurrent layers~\cite{argun2021classification, li2021wavenet}, graph neural networks~\cite{verdier2021learning}, Bayesian inference~\cite{park2021bayesian}, or extreme learning machines~\cite{manzo2021extreme}. All these methods are trained in a supervised scheme, which means that the machine learns to characterize the data by training with human-labelled data. While for some applications,  such as parameter estimation, this approach is completely valid and advantageous, for others it creates an inherent problem that may deeply affect our understanding of the data analyzed.

An example of such inherent problem is the discrimination of the diffusion model of the particles, which has been usually stated as a ML classification problem~\cite{dosset2016automatic,kowalek2019classification, munoz2020single}. When classifying between diffusion \textit{modes} (i.e. normal, anomalous, confined or directed diffusion), the problem is simplified, as they probably contain all possible kinds of diffusion. However, a different problem is to classify between diffusion \textit{models}, such as e.g. fractional Brownian motion (FBM) or continuous time random walk (CTRW) (see Section~\ref{sec:data} for further details on these). In this case, it is very possible that the actual model of the data analyzed is not contained in the training set, or may even be a combination of these. Usual supervised approaches lack the tools to discriminate these cases and they associate them with the most resembling model~\cite{munoz2020single}.

In this work, we explore the use of unsupervised techniques for single trajectory characterization and show that they can indeed be used to analyze anomalous diffusion data coming both from simulations and various experimental setups. In particular, we focus on the use of neural network autoencoders, a typical ML architecture for unsupervised and semi-supervised approaches. Our aim is: first, to investigate if unsupervised learning yields the same results of supervised techniques previously studied; second, to study the use of unsupervised approaches to understand and compare different anomalous diffusion models beyond supervised methods; third, to explore the possibility of finding new physical processes related to anomalous diffusion by means of autoencoders.

The work is organized as follows: in Section~\ref{sec:methods} we present the unsupervised approach of use, namely convolutional autoencoders (Section~\ref{sec:ae}) and anomaly detection (Section~\ref{sec:anomaly}). Moreover, in Section~\ref{sec:data} we present the anomalous data used, as well as further details on the anomalous diffusion models considered. Then, the results of the paper are presented in Section~\ref{sec:results}. First, we explore the feasibility of anomaly detection in simulated data (Section~\ref{sec:simulated}), both for model discrimination (Section~\ref{sec:models}) and parameter estimation (Section~\ref{sec:parameter}, but also to characterize novel types of diffusion, showcased in this case by the combination of existing models. (Section~\ref{sec:composite}) We then apply the knowledge gained to analyze experimental data arising from SPT trajectories (Section~\ref{sec:experiments}).

\section{Methods}
\label{sec:methods}

\begin{figure*}
    \centering\includegraphics[width=0.8\textwidth]{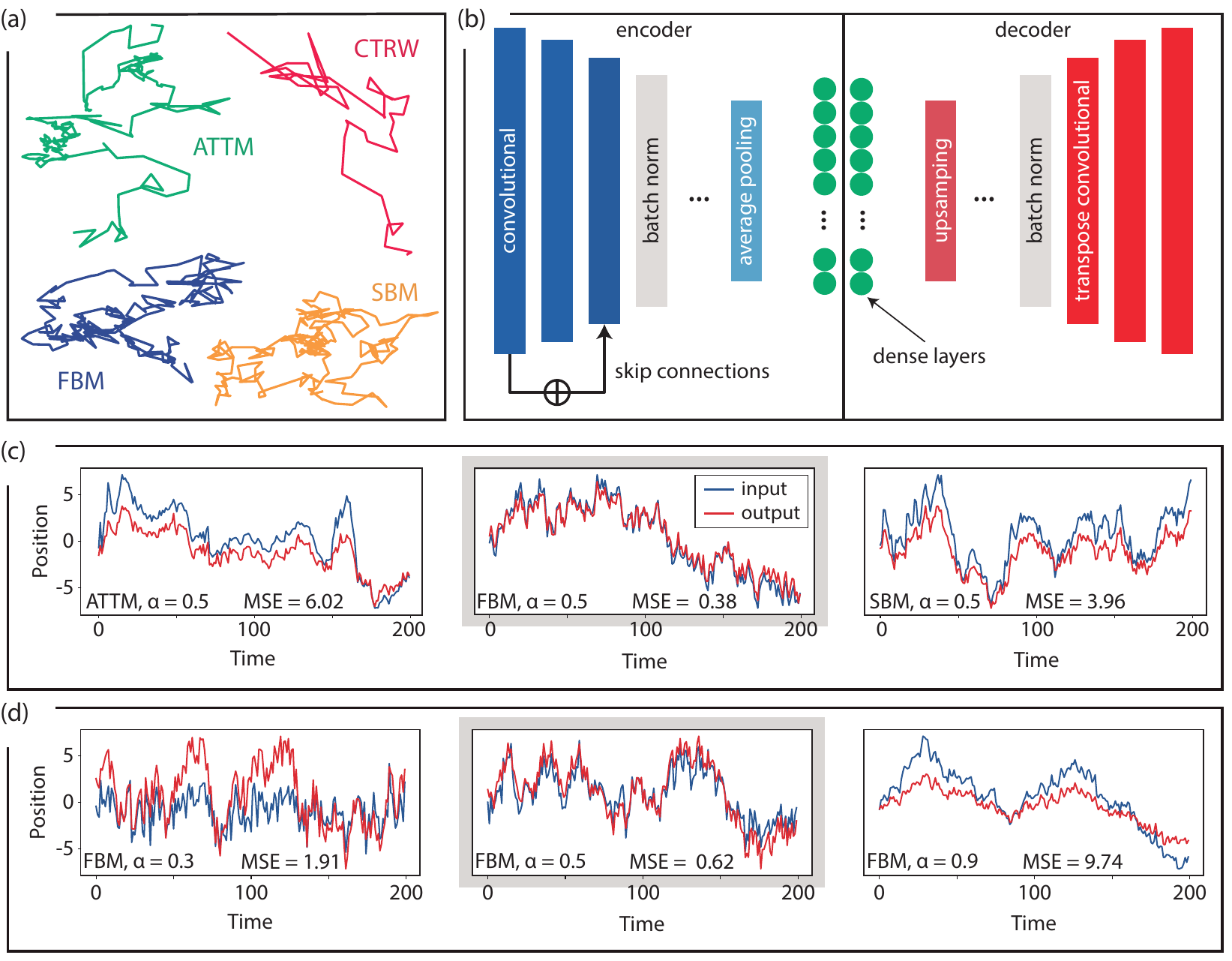}
    \caption{
    (a) Representative two dimensional trajectories of the diffusion models considered in this work. All trajectories have anomalous exponent $\alpha = 0.7$.
    (b) Schematic representation of the autoencoders used, showcasing some of the layers composing them, as well as the \textit{skip connections} feature.
    (c) Input and output of an autoencoder trained with FBM at $\alpha=0.5$ (highlighted in gray), predicting trajectories of ATTM, FBM and SBM with $\alpha=0.5$, respectively. 
    (d) Input and output of the same autoencoder, now with trajectories of FBM at three different exponents, $\alpha = [0.3, 0.5, 0.9]$.
    In both (c-d), the best reconstruction (lowest MSE) occurs for trajectories of the model/exponent with which the autoencoder was trained.}
    \label{fig:scheme}
\end{figure*}

\subsection{Autoencoders}
\label{sec:ae}

Autoencoders (AE) are artificial neural networks that learn efficient encodings of high-dimensional data. They consist of two parts: an encoder that compresses the input data into a low-dimensional embedding, referred to as the \textit{latent space}, and a decoder that recovers the original input from the compressed representation. Ideally, the  output of the decoder is equal to the input of the encoder. However, due to the compression of information happening in the latent space, such an ideal scenario is often not found. Because of such compression, in order to improve the similarity between reconstructions and inputs, the AE has to learn features of the dataset during the training. Such features will then be used to plug the information gaps caused by the encoder's compression.

In our particular problem, the training data consists of a set of simulated anomalous diffusion trajectories $X = \{x^{(1)},x^{(2)},\dots, x^{(n)}\}$, where each training sample $x_i$ consist of a one dimensional single trajectory with $T$ timesteps, $x^{(i)} = [x^{(i)}_1, x_2^{(i)} , ..., x_T^{(i)}  ]$, with $x\in \mathbb{R}$. A typical loss function for the reconstruction of such data is the mean squared error (MSE), which for a single trajectory  reads
\begin{equation}
\mathrm{MSE} = \frac{1}{T}\sum_{t=1}^T (f(x^{(i)}_t) -x^{(i)}_t)^2,
\label{eq:MSE}
\end{equation}
where $f(x^{(i)})$ is the AE output for the input trajectory $x^{(i)}$.

The AE used consists of a stack of convolutional and fully-connected layers. Moreover, we consider features such as batch normalization, as well as a global average pooling and upsampling before and after the latent space. An schematic representation of such AE is presented in Fig.~\ref{fig:scheme}(b). In this scheme, connections exist between subsequent layers, as typically occurs in most neural networks architectures, but extra connections are created between not subsequent ones. Such a special feature is often referred to as \textit{skip connections}, and is widely used in residual neural networks~\cite{he2016deep}. Heuristically, we find that the use of such connections boosts the power of the AE, increasing vastly its accuracy. Moreover, due to the stochastic nature of the trajectories, any decrease in the dimensionality of the trajectory (i.e. to the size of the latent space) will cause the loss of necessary information for its reconstruction. In that sense, the reconstruction error will always have a lower bound related to the size of the latent space~\cite{epstein2019generalization}. In our particular implementation, the latent space consists of a fully connected layer whose size will depend on the length of the trajectories $T$. A detailed description of the architecture is presented in Appendix~\ref{app:ae_architecture} as well as in the repository of Ref.~\cite{github}.

Importantly, in order to avoid the effect of overfitting of the AEs, and to ensure that their predictions are related to the learnt features and not the memorization of the data, all the results presented in this manuscript correspond to predictions over sets of trajectories not used in the training.

\subsection{Anomaly detection}
\label{sec:anomaly}

Anomaly detection refers to the problem of detecting instances or samples from a certain dataset that differ in some sort from the rest. In other words, the goal is to detect outliers, a.k.a. anomalies, which while having similarities with the rest of the dataset, also have special features that crucially make them different. In recent years, machine learning methods have shown great capabilities in dealing with such problem~\cite{chalapathy2019deep}, with interesting application in Physics~\cite{kottmann2020unsupervised}.

Particularly, AEs offer a powerful and versatile architecture for anomaly detection while relying on unsupervised or semi-supervised learning~\cite{zhou2017anomaly}. As commented in the previous section, the goal of an AE is to reconstruct an input instance with minimal error. In this sense, anomalies can be detected by setting a threshold to this error, in such a way that any value above indicates that the input reconstructed is indeed an anomaly of the dataset. In this particular work, the prediction error is given by the MSE, Eq.~\eqref{eq:MSE}. Heuristically, we find that such a metric, in conjunction with the data preprocessing presented in Section~\ref{sec:data}, yields good results and suffices for the task at hand. Nevertheless, other interesting approaches, based on the reconstruction probabilities~\cite{an2015variational} rather than absolute values are left to be explored in future works. 

\begin{figure}
    \centering\includegraphics[width=\columnwidth]{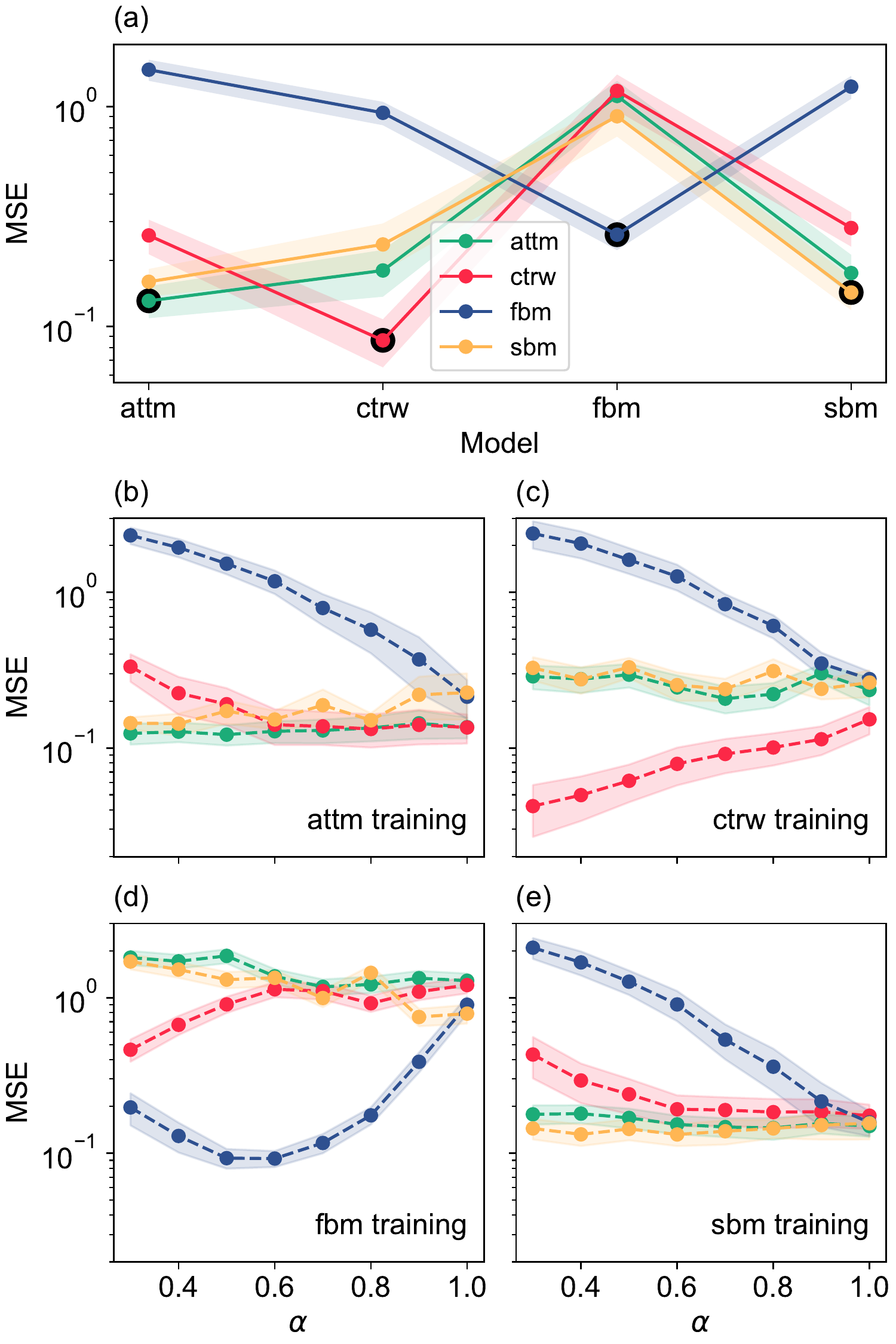}
    \caption{
    (a) MSE vs. the diffusion model of the input trajectories. Each line corresponds to an AE trained with the indicated model.
    (b-e) MSE vs. anomalous exponent $\alpha$. Each panel corresponds to an AE trained with the indicated diffusion model, while the color of the lines indicate the model of the input trajectories, following the legend of (a).}
    \label{fig:one_vs_rest}
\end{figure}

\subsection{Anomalous diffusion data}
\label{sec:data}

As commented, the datasets considered consist of trajectories arising from different anomalous diffusion models. In particular, we consider four different models, with which we aim to cover a large variety of real-world phenomena: annealed transient time motion (ATTM)~\cite{massignan2014nonergodic}, continuous-time random walk (CTRW)~\cite{scher1975anomalous}, fractional Brownian motion (FBM)~\cite{mandelbrot1968fractional}, L\'{e}vy walk (LW)~\cite{klafter1994levy},  and scaled Brownian motion (SBM)~\cite{lim2002self}. Moreover, these models allow us to explore the whole range of anomalous diffusion exponents $\alpha$: ATTM and CTRW are subdiffusive models ($\alpha\leq 1$) and FBM and SBM cover the whole range (i.e. $0<1<2$). A complete review on the properties of each model can be found in Ref.~\cite{metzler2014anomalous, munoz2021objective}.

The trajectories that we analyzed here were generated via the ANDI dataset python package~\cite{munoz2021objective}. Moreover, each trajectory is normalized such that the standard deviation of their displacements is equal to 1. This ensures that all trajectories have a similar diffusion coefficient $D\approx 1$, but the anomalous exponent is unaltered. Moreover,  to avoid very slow and immobile trajectories, we will restrict the dataset to $\alpha >= 0.1$. Finally, all the results presented in this work corresponds to AE trained with trajectories of only 20 time steps, if nothing different is stated. While similar results are found for longer trajectories, our goal is to showcase the applicability of the method for experimental scenarios, where such short trajectories are often the only ones accessible.

In Fig.~\ref{fig:scheme}(c-d) we briefly showcase the expected results when dealing with anomaly detection with anomalous diffusion trajectories. Depending on the features of the input data, the AE will do a better or worst reconstruction. For instance, in Fig.~\ref{fig:scheme}(c) we showcase that an AE trained with FBM yields better results, seen both visually and with the values of the MSE, when reconstructing trajectories of this same model (highlighted in gray), rather that trajectories of ATTM (left) or SBM (right). Similarly, in Fig.~\ref{fig:scheme}(d),  we showcase that the MSE is lower when dealing with trajectories of the same exponent the AE was trained with. In the next sections we will deeply analyze this behaviour and understand the power and limitations of the proposed anomaly detection method.

\section{Results}
\label{sec:results}

We present here the results on applying anomaly detection to anomalous diffusion trajectories. In particular, we focus here on two important aspects of the trajectories: their diffusion model and their anomalous exponents. Our aim is to show that an autoencoder can learn to reconstruct trajectories arising from different anomalous diffusion conditions. More importantly, after training, the AE must be capable of reconstructing new, unseen trajectories. As commented previously, the reconstruction error will then be used as a measure of how related the new trajectories are w.r.t. the ones used for training. In order to benchmark the method in a controlled scenario, we will first present the results on simulated trajectories, and then show its applicability to data from various experimental conditions.

\subsection{Simulated trajectories}
\label{sec:simulated}

\subsubsection{Detecting changes of diffusion model}
\label{sec:models}
Our first result relates to the ability of the method to differentiate between distinct diffusion models. We start with the simplest scenario, in which an AE is trained with a single diffusion model.
Then, we compare its error (as given by the MSE, Eq.~\eqref{eq:MSE}) when reconstructing trajectories of other models. In Fig.~\ref{fig:one_vs_rest}(a) we show the results when training AEs with subdiffusive trajectories with the models indicated in the legend. Each line corresponds to the averaged error over ten different AE and the shaded area represents their respective standard deviations. Then, we reconstruct trajectories with each of the AE for different models, as shown in the x-axis. Importantly, in all cases, the error of the AE is minimal in the model it was trained with (highlighted with a black rounded circle). Moreover, we can already infer, based on the reconstruction error, similarities between some of the models. For instance, the AEs trained with ATTM and SBM trajectories share a very similar behaviour. This is to be expected as they are the most resembling from the pool of chosen models. Indeed, both are so-called random diffusivity models, i.e. showing Brownian-like diffusion but with non-trivial changes of the diffusion coefficient. We note that, even in the case of supervised and model specific algorithms, their discrimination is very challenging~\cite{munoz2021objective}. In the case of the FBM trajectories, we see that the error made by AEs is an order of magnitude larger than the rest of the models (note the logarithmic scale). This showcases the strong differences between FBM and the rest of models, being the former the only ergodic model, as well as the only with correlated displacements.

To further understand the presented results, we study how the MSE behaves for the different models at different anomalous exponents $\alpha$, as shown in Fig.~\ref{fig:one_vs_rest}(b-e). Each panel corresponds to an AE trained with the indicated model, while each line corresponds to trajectories of the different models (as indicated by the legend in Fig.~\ref{fig:one_vs_rest}(a)). These AE were trained in the whole range of $\alpha$ showcased. We see here that when $\alpha \to 1$, trajectories of all models collapse to the same MSE, no matter the AE used. This is an important result, as even if their microscopical behaviour may be different, all models converge to the Brownian motion universality class at $\alpha=1$ (up to logarithmic corrections~\cite{klafter2011first}). Hence, the AEs are able to learn reliable features from the anomalous diffusion trajectories and not just trivial patterns.

Such results also allow us to further compare the different models considered. Again, we see that the non-ergodic models tend to share very similar MSE. For instance, the AEs trained with ATTM (Fig.~\ref{fig:one_vs_rest}(b)) and SBM (Fig.~\ref{fig:one_vs_rest}(e) are able to correctly reconstruct all models but FBM. However, for CTRW trajectories the best results are found when trained with the latter model (Fig.~\ref{fig:one_vs_rest}(d), as the characteristic trapping times of CTRW may be difficult to reproduce if not learnt in the training. As previously seen, the error made by the FBM trained AE shows errors an order of magnitude larger when reconstructing other models, which nevertheless compare to the error of FBM at $\alpha=1$.

\subsubsection{Detecting parameter changes within a diffusion model}
\label{sec:parameter}
\begin{figure}
    \centering\includegraphics[width=\columnwidth]{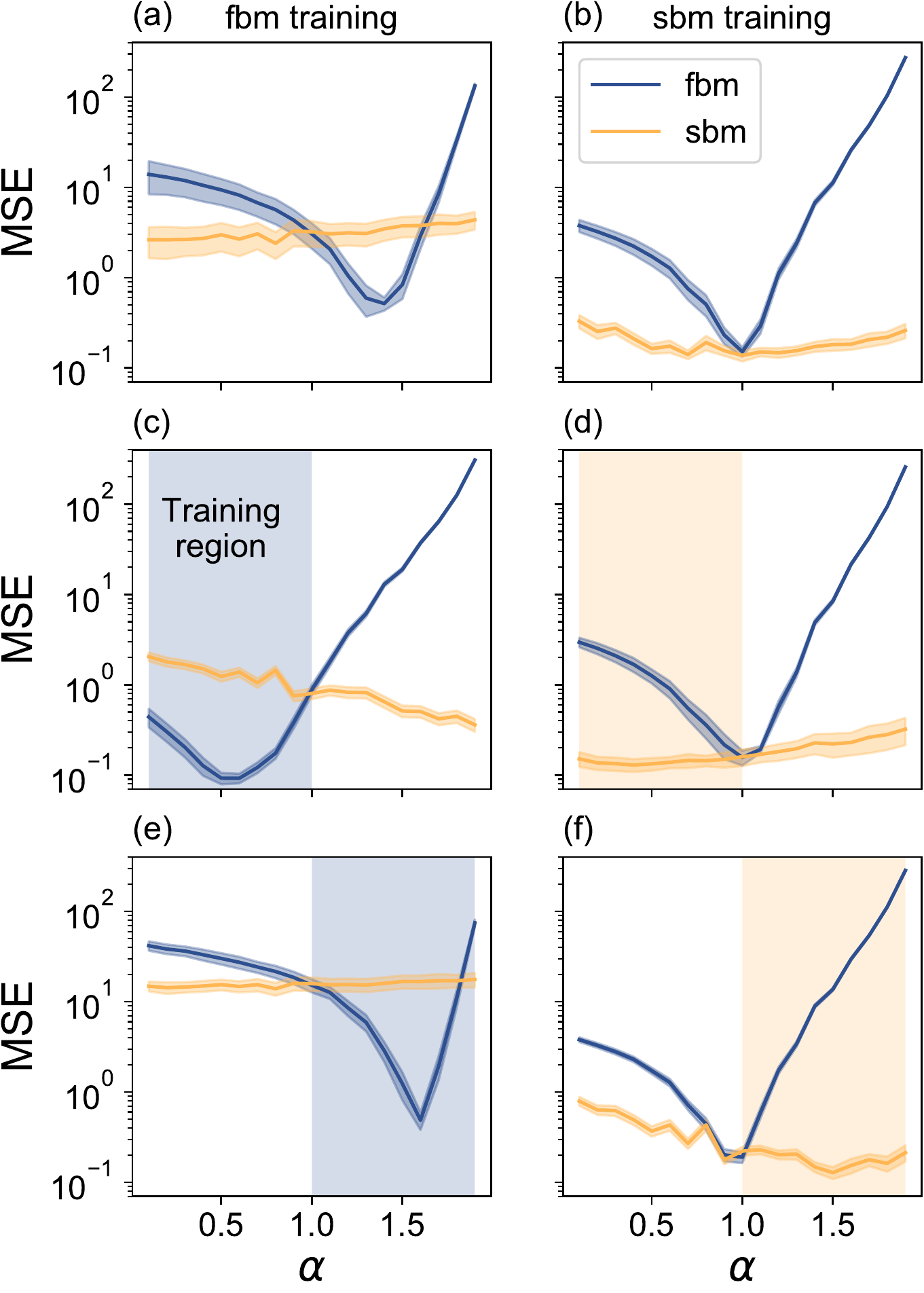}
    \caption{MSE vs. anomalous exponent $\alpha$ for AEs trained with FBM (left) or SBM (right) trajectories in different ranges of $\alpha$ (as indicated by the shaded regions):
    (a-b) $\alpha \in [0.1, 2)$;
    (c-d) $\alpha \in [0.1, 1]$;
    (e-f) $\alpha \in [1, 2)$}
    \label{fig:fbm_vs_sbm}
\end{figure}

While the AEs trained with ATTM, CTRW and SBM show an almost constant MSE at all $\alpha$ for their own model, the one trained with FBM tends to minimize the error for a subset of exponents. Taking advantage that both FBM and SBM are able to generate both subdiffusive and superdiffusive trajectories, we trained AEs in $\alpha\in \left [ 0, 2 \right]$. In Fig.~\ref{fig:fbm_vs_sbm}(a-b) we show such results for FBM and SBM, respectively. The line color indicates the model of the input trajectories. Interestingly, the FBM trained AE shows the same MSE behaviour stated above, but now the minimal MSE occurs at a different exponent. Surprisingly, in the subdiffusive range, the MSE is lower for SBM  than for FBM (and also in the highest part of the superdiffusive range). This does not occur in the SBM trained AE, where the MSE is almost constant for all $\alpha$ and is always minimal for the model it was trained with. Again, in this case, the minimal error for FBM trajectories is found at $\alpha = 1$.

The FBM model offers an interesting playground to study the suitability of anomaly detection in anomalous diffusion. Within the same diffusion model, FBM shows two completely different behaviours, depending on the value of $\alpha$. For $0<\alpha<1$, FBM is anti-persistent, hence showing negative correlations, while it is persistent and positively correlated for $1<\alpha<2$. This makes trajectories for different $\alpha$ very different from each other, hence the difficulty of an AE to correctly reconstruct trajectories at all $\alpha$. Such behaviour is highlighted in Fig.~\ref{fig:fbm_vs_sbm}(b-e), where AEs were trained with FBM and SBM in the subdiffusive (b-c) and superdiffusive (d-e) ranges. While the information learnt by the SBM trained AE in any of these two ranges is sufficient to correctly reconstruct trajectories for all $\alpha$, the one trained with FBM only yields low errors for the range it was trained on.

In order to understand further this behaviour, we analyze the reconstructed trajectories output by the AE trained with two models, FBM and CTRW. We note that the results found for CTRW are analog to the ones found with SBM and ATTM. We proceed to input trajectories to the AE with different $\alpha$. Then, we calculate the anomalous exponent of the output trajectories by either fitting an ensemble average of the MSD (for CTRW) or by performing a time averaged MSD (for the FBM). A linear fit in the logarithmic scale outputs the anomalous exponent $\alpha_f \approx \alpha$.

\begin{figure}
    \centering\includegraphics[width=\columnwidth]{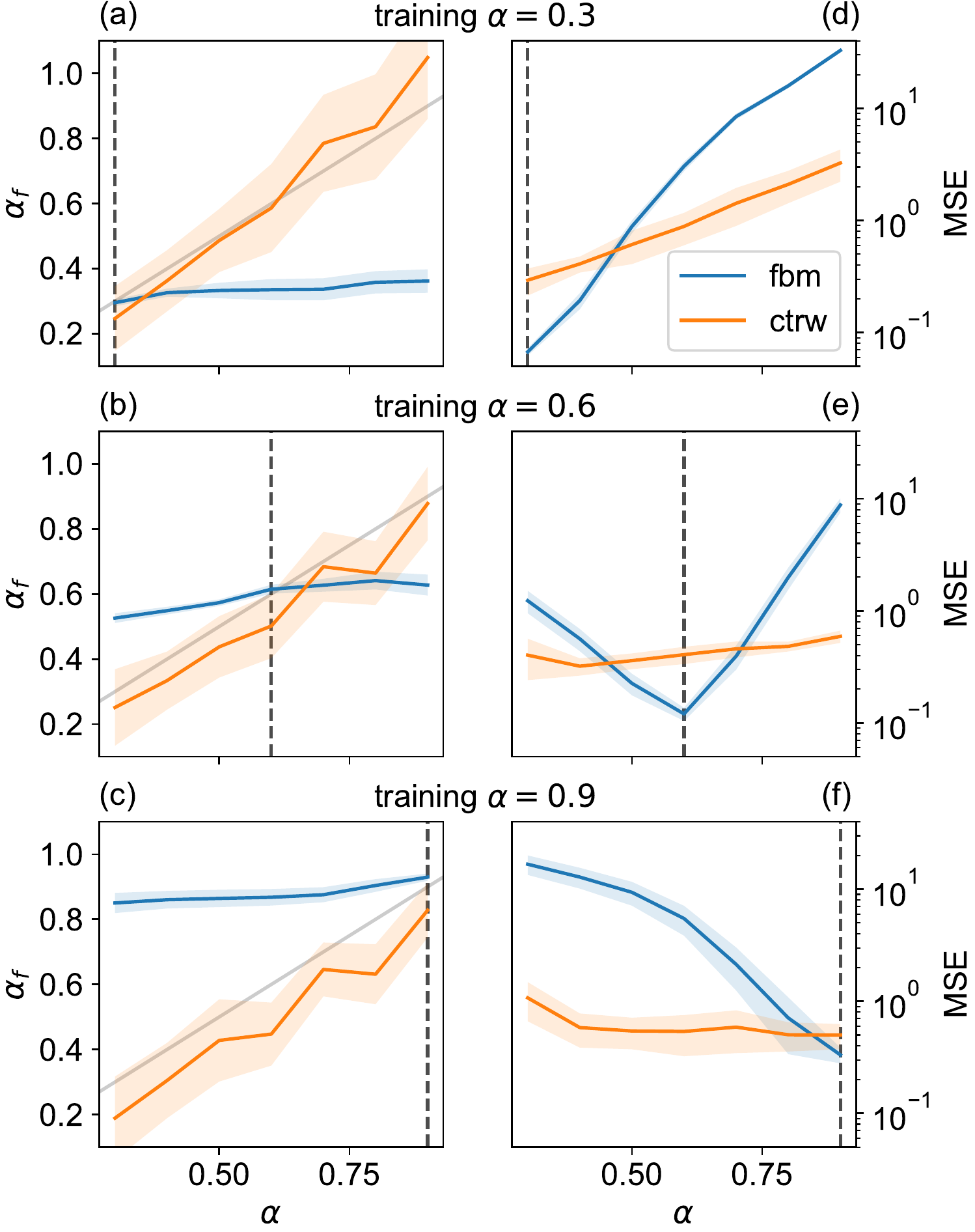}
    \caption{
    (a-c) Fitted anomalous exponent $\alpha_f$ of the reconstructed trajectories vs. the ground-truth exponent $\alpha$ of the input trajectory. Gray line corresponds to $\alpha_f = \alpha$.
    (d-f) MSE vs. anomalous exponent $\alpha$. For all plots, we consider AEs trained with FBM (blue) or ATTM (orange). Each row corresponds to AEs trained at different $\alpha$, indicated in the title and the vertical dashed line.}
    \label{fig:alpha_effect}
\end{figure}

We present results for AEs trained at three different $\alpha$, as shown in Fig.~\ref{fig:alpha_effect}. For this test, we consider trajectories only of the model the AE was trained with (i.e. FBM and CTRW in this case). Moreover, to allow for more accurate estimation of $\alpha_f$ via the ensemble and time averaged MSD, the AEs trained with trajectories of $T=200$ time steps. In the left column,  we show the fitted $\alpha_f$ w.r.t. to the exponent of the input trajectories. In the right column we show the reconstruction error for these same AEs, again as a function of the exponent of the input trajectories. In all cases, for the CTRW trained AE (orange), no matter the exponent it was trained with, the output trajectories will maintain the exponent of the input (orange line in Fig.~\ref{fig:alpha_effect}(a-c)). Surprisingly, even if the MSE increases for $\alpha$s different of the one used in the training (see e.g. Fig.~\ref{fig:alpha_effect}(d)), the autoencoder is able to reproduce a key feature of the CTRW trajectories as is the anomalous exponent. Oppositely, the outputs of the AE trained with FBM are transformed such that their $\alpha_f$ equals the one used for training (blue line in Fig.~\ref{fig:alpha_effect}(a-c)). This hence causes that the MSE grows significantly when dealing with trajectories of different $\alpha$, as we saw already in Fig.~\ref{fig:fbm_vs_sbm} and again highlighted in the right column of Fig.~\ref{fig:alpha_effect}.

We note here that while both CTRW and FBM trajectories can be characterized by the anomalous exponent, the source of anomalous diffusion complete differs from one another. In the case of the CTRW, the exponent is related to the distribution of waiting or trapping times of the particle. However, in the case of FBM, the exponent is connected to the displacement's correlations. The previous results showcase that the later phenomena is a crucial feature for the FBM trained AE, as once a certain correlation is learnt (i.e. one $\alpha$ is learnt), all the reconstructed trajectories share that same correlation. However, in the case of CTRW, the AE is able to reconstruct trajectories without 'loosing' their characteristic exponent, even with different $\alpha$ of the one used from training. The reason of this difference is still unknown and further exploration of such phenomena is currently ongoing with more sophisticated AE.

\subsubsection{Detecting mixed diffusion models}
\label{sec:composite}

To further understand the extent of use of unsupervised learning in anomalous diffusion data, we test the method in trajectories arising from the composition of different diffusion models. Indeed, it is expected that the diffusion of particles in complex environments, such as biological systems, is not described by a single model but the intertwine between two or more. For instance, in Refs.~\cite{ weigel2011ergodic, tabei2013intracellular} the diffusion of particles through the cell membrane was described by a particle performing a FBM in the presence of a fractal structure and traps, respectively. The latter phenomena can be associated to a CTRW, hence leading to the intertwined diffusion of CTRW and FBM. In the small temporal scale, the dominant model would be FBM. However, in a bigger picture, traces of CTRW such as a broad distribution of trapping times will arise. Recently, other kind of diffusion model compositions have been studied, in which particles switch between different diffusion states~\cite{janczura2021identifying, sabri2020elucidating}. Nevertheless, we believe that such problem may be better solved with via segmentation algorithms~\cite{argun2021classification, munoz2021objective} and focus on the former subordinated phenomena.

\begin{figure}
    \centering\includegraphics[width=\columnwidth]{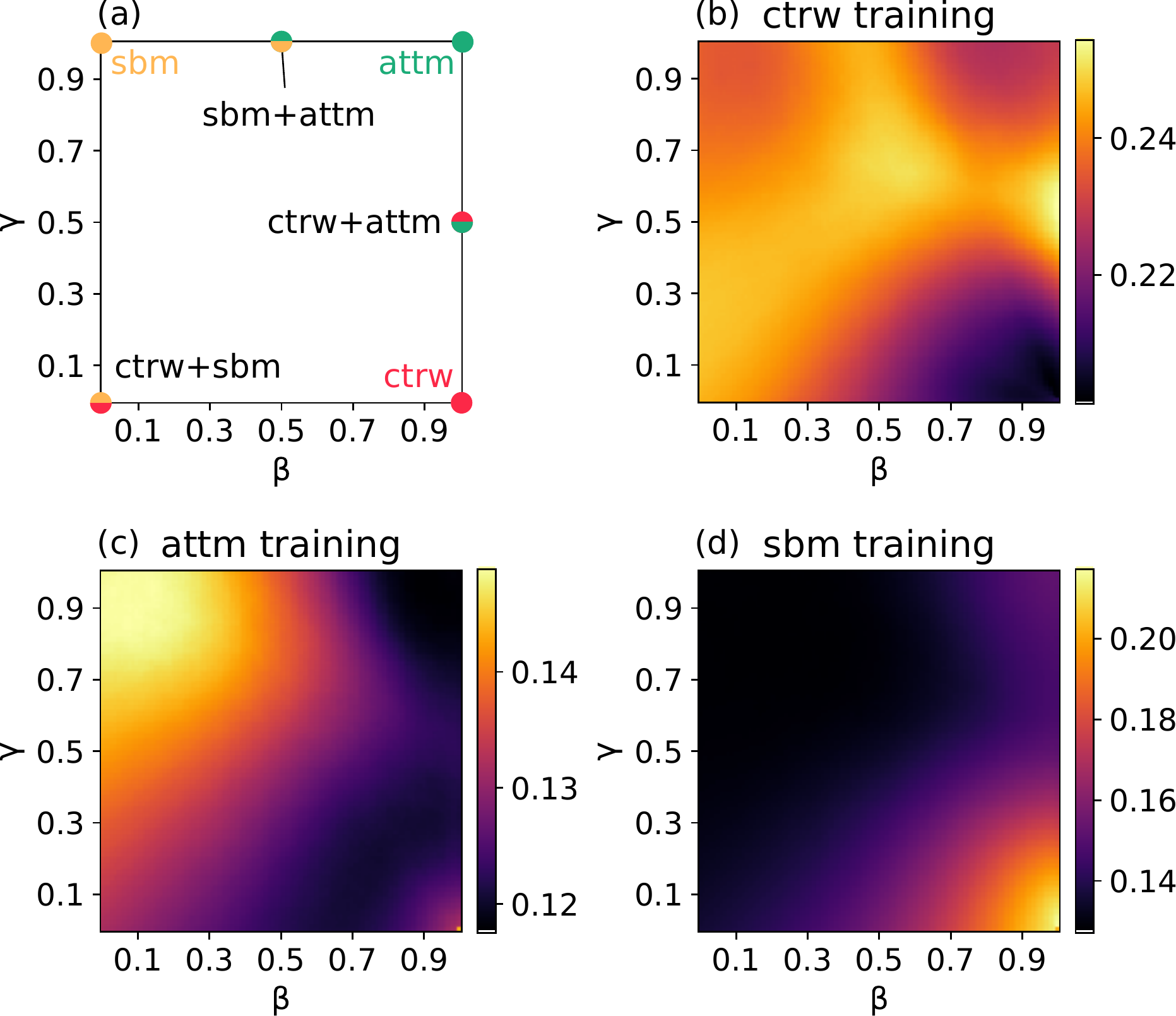}
    \caption{
    (a) Schematic representation of the diffusion models governing a trajectory with certain $\gamma$ and $\beta$ as given by Eq.~\eqref{eq:composite}.
    (c-d) MSE as a function of the trajectories $\gamma$ and $\beta$ for AE trained with CTRW, ATTM, and SBM trajectories in the subdiffusive range $\alpha\in[0.1,1]$ and $T=20$. Each point averages the results of 10 AE over 5000 trajectories.}
    \label{fig:composite}
\end{figure}

In order to simplify the present example, we consider trajectories whose displacements $\Delta^i_{\gamma,\beta}$ are the weighted sum of displacements of three different diffusion models:
\begin{equation}
\label{eq:composite}
\Delta_{\gamma,\beta}^i =
\gamma \beta \Delta_{\rm{model}_1}^i +
(1-\gamma) \Delta_{\rm{model}_2}^i +
(1-\beta) \Delta_{\rm{model}_3}^i,
\end{equation}
where $\Delta^i_{\rm{model}_j}$ refers to the $i$-th displacement of a trajectory of the model $j$. We note that the displacements used to generate a composite trajectory come from a single trajectory of each model, hence maintaining their properties (e.g. correlations and ageing effects). Then, the parameters $\gamma$ and $\beta$ allow us to change the dominant model, creating the phase space schematically represented in Fig.~\ref{fig:composite}(a).

We start our analysis by performing anomaly detection of trajectories arising from Eq.~\eqref{eq:composite} with AE trained with \textit{pure} models, as shown in Fig.~\ref{fig:composite}(b-d). As expected, the loss is minimized in the corner associated with the diffusion model used for training. Interestingly, the AE trained with CTRW favours trajectories of pure models, while the MSE is maximal in the regions where the models are equally weighted. In the case of the ATTM trained AE, the MSE increases when approaching the SBM corner and in the case of pure CTRW. Such a result reproduces what is expected from a theoretical point of view: the lower right sector of the phase diagram can be understood as a noisy CTRW model, which is indeed very difficult to differentiate from an ATTM trajectory in the trajectory length of work ($T=20$). Last, the SBM trained AE seems to correctly reconstruct a much bigger region of the phase space, while favouring ATTM over CTRW. We note that the results found in the corners of the phase space are completely analogous to the ones in Fig.~\ref{fig:one_vs_rest}. As expected, most of the conclusions drawn here are also similar to the one discussed in Section~\ref{sec:models}. However, this approach also for a much more refined analysis of the trajectories, allowing for a deeper level of understanding and the possibility of finding novel types of diffusion, as will be discussed in Section~\ref{sec:experiments}.

The previous results allow us to validate the use of the composite models and their analysis with AE trained with single models. A different approach is to consider autoencoders trained with trajectories of different $\gamma$ and $\beta$. Then, performing anomaly detection with such AE will allow us to characterize the model of the input trajectories by finding the $\gamma$ and $\beta$ with minimal MSE. Such approach has no particular interest applied to the considered simulated trajectories, and will hence be directly applied to experimental datasets with unknown models.

\subsection{Experimental trajectories}
\label{sec:experiments}

The previous results showcased the validity of the presented method to detect variations and outliers in simulated data. However, as commented previously, the main utility of these techniques relies on their implementation in experimental data, as they provide a fast and accurate method to analyze anomalous diffusion trajectories. To showcase that unsupervised methods can be used for that aim, we present here results on the use of anomaly detection in trajectories from various single particle tracking experiments. While the focus here is in biophysical problems, as stated in the Introduction, these techniques are applicable at any scale or system where stochastic trajectories arise.

\begin{figure}
    \centering\includegraphics[width=\columnwidth]{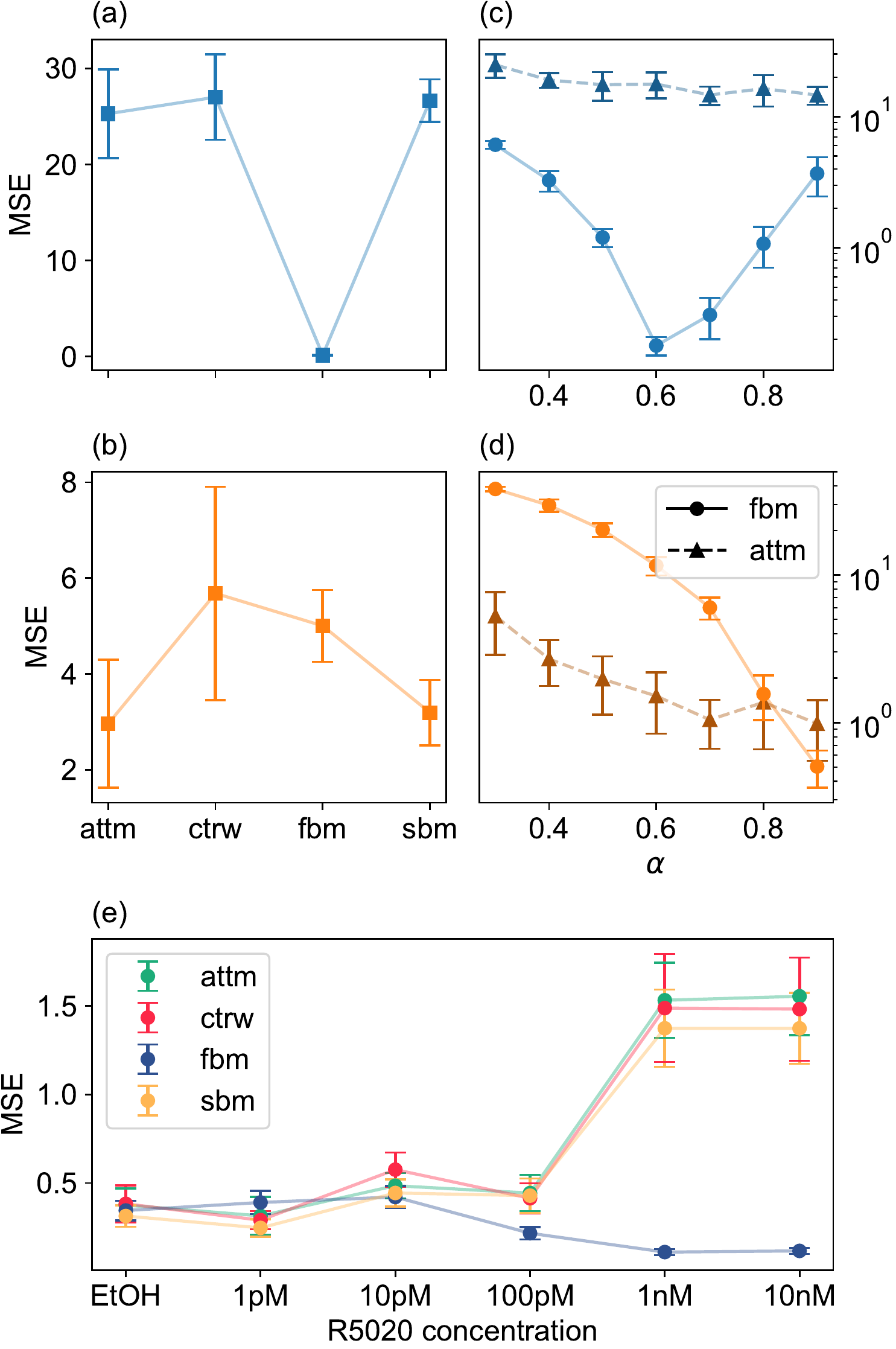}
    \caption{Anomaly detection in three experimental datasets. Each row corresponds to Refs.~\cite{stadler2017non,krapf2019spectral}, Ref.~\cite{manzo2015weak} and Ref.~\cite{munoz2020phase}, respectively.
    (a-b) MSE vs. the diffusion model used to train the AE.
    (c-d) MSE. vs. the anomalous exponent $\alpha$ for AEs trained with FBM (circles, solid line) and ATTM (triangles, dashed line).
    (e) MSE for four AEs trained with the indicated diffusion model at different R5020 hormone concentration.}
    \label{fig:experiments}
\end{figure}

Here we consider trajectories from three different experiments and present their results in each row of Fig.~\ref{fig:experiments}. The trajectories of the two first experiments were cropped to have 200 time steps while for the third trajectories have only 20 time steps. Hence, different AEs will be used for each length.  First, we analyze the trajectories of telomeres in the nucleus of mammalian cells~\cite{stadler2017non, krapf2019spectral}. In these two references, the authors showed that the trajectories are well described by a FBM model, with a crossover in the anomalous exponent from $\alpha \sim 0.5$ at short times to $\alpha \sim 1$ for longer times. The second dataset corresponds to the trajectories of receptors diffusing in the plasma membrane of mammalian cells~\cite{manzo2015weak}. These trajectories have been shown to be consistent with the ATTM and $\alpha\sim 0.8$.

Before presenting the third dataset, we focus on the results found for the latter datasets. We start by inputting the trajectories to AEs trained with different diffusion models. In the case of the telomeres (Fig.~\ref{fig:experiments}(a)), the MSE is largely reduced when reconstructing trajectories with a FBM trained AE, hence coinciding with the expected results. For the diffusion of receptors (Fig.~\ref{fig:experiments}(b)), the MSE is minimal for ATTM, but also very close to the SBM one. While these results are obviously not conclusive, they already hint at important information and are indeed very well related to what we showed for simulated trajectories above (i.e. shared behaviour between SBM and ATTM). Moreover, this is also consistent with previous ML supervised approaches (see Fig.5(c) of Ref.~\cite{munoz2021objective}).

When performing anomaly detection with different anomalous exponents, the results on each dataset differ, as the assigned diffusion models are also different. For both datasets, we show in Fig.~\ref{fig:experiments}(c-d) the MSE w.r.t. to the anomalous exponent $\alpha$ of the training dataset, for AEs trained with FBM (solid line, circles) and ATTM (dashed line, triangles). For the diffusion of telomeres, there is a sharp decrease of the MSE for the FBM trained AE (Fig~\ref{fig:experiments}(c), circles), indicating an expected exponent of $\alpha = 0.6$, consistent with previous predictions. For the receptors' trajectories, this same AE has a minimum at $\alpha = 1$, an expected result as our previous analysis shows that these trajectories do not diffuse as FBM. Hence, the MSE is expected to be minimized at $\alpha=1$. This result is analogous to what was found in Section~\ref{sec:models} and Fig.~\ref{fig:one_vs_rest}(b-e)). Inputting the receptors' trajectories to an ATTM trained AE, the minimisation of the AE occurs in $\alpha \in [0.7,1]$. While these results are in accordance with previous works, we note that AE trained with ATTM at different $\alpha$ are able to correctly reconstruct trajectories from the rest of the range, as shown in Fig.~\ref{fig:fbm_vs_sbm}(d-f). We note again that, while the exploration of diffusion models by means of AE gives rise to great opportunities, we believe that the extraction of diffusion parameters, such as in the case the anomalous exponent, will always be better with supervised techniques.

At last, we focus on the motion of the Progesterone receptor as it diffuses in the cell nucleus, under the presence of the R5020 hormone~\cite{munoz2020phase}. This system shows liquid-liquid phase separation (LLPS) above a certain hormone concentration. Interestingly, two populations coexist, one where particles diffuse freely through the cell membrane, following ATTM and $\alpha \sim 0.8$ and a second one containing the particles interacting with the chromatin and forming phase separated aggregates, following FBM and $\alpha \sim 0.4$. As the hormone concentration increases, more and more particles interact with the chromatin and hence the FBM population grows.

Importantly, such result is also found by means of anomaly detection. In Fig.~\ref{fig:experiments}(e) we show the MSE for four AEs, trained with the diffusion model indicated in the legend. While the MSE is similar for all AEs at low hormone concentrations, we see a sharp increase above 10 nM. This is consistent with the results shown in Ref.~\cite{munoz2020phase} and corresponds to the point in which most of the particles in the system phase-separate. This makes the FBM population increase drastically. Hence, the AE trained with such a model decreases its error, while the rest see a big increase. At low concentrations, the ATTM and FBM populations are balanced, hence the errors for ATTM and FBM are of the same order. Interestingly, the AEs trained with CTRW and SBM show a similar behaviour as the ATTM one, following what we already saw in simulated trajectories (see Fig.~\ref{fig:one_vs_rest}). We note here an important remark: the trajectories of this last experiment are extremely short (only 20 time steps). This makes any proper statistical analysis extremely challenging and even impossible. While the presented results may not be conclusive in differentiating the non-ergodic models, it already hints very important information that was inaccessible to non machine learning approaches. Moreover, being this an unsupervised method, the AE’s predictions are based solely on the features learnt and not in the need of minimizing a classification based metric, as may happen in supervised learning.

\begin{figure}
    \centering\includegraphics[width=\columnwidth]{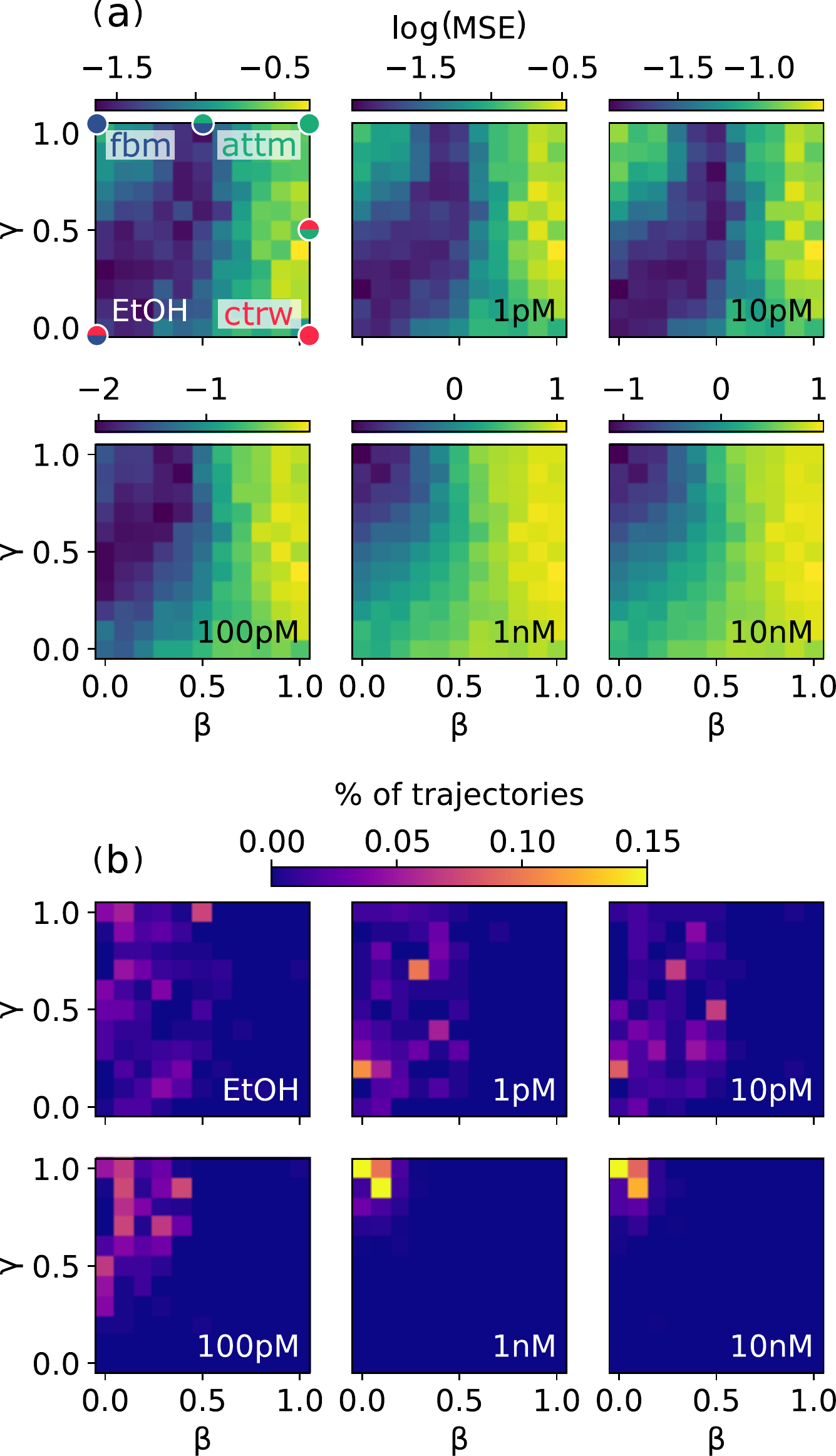}
    \caption{(a) MSE of the experimental dataset of Ref.~\cite{munoz2020phase} for AE trained with composite models (Eq.~\eqref{eq:composite}) made of CTRW, ATTM and FBM at various $\gamma$ and $\beta$.
    (b) Percentage of trajectories whose minimal MSE is found for the AE trained at the given $\gamma$ and $\beta$. Each point corresponds to the average over 7 different autoencoders.}
    \label{fig:composite_PR}
\end{figure}

To further investigate the diffusion of the PR, we use a similar method as the one presented in Section~\ref{sec:composite}. In this case, we will train AE with trajectories at different $\gamma$ and $\beta$. For this particular example, we consider CTRW, ATTM and FBM as model 1,2,3, respectively (see Eq.~\eqref{eq:composite}). This yields the phase space schematically represented by the points and labels of the first panel of Fig.~\ref{fig:composite_PR}(a). First, we calculate the MSE over the set of trajectories with AE trained at different $\gamma$ and $\beta$. At low hormone concentrations, the minimal MSE is found with AE trained with mixed models, all with the presence of FBM. Such result is consistent to what was found in Fig.~\ref{fig:experiments}(e), where all models shared similar MSE. Interestingly, not any mix of models yields low MSE, favouring mixes with FBM. Then, at higher hormone concentrations, the lowest MSE is found with AE trained with trajectories with higher FBM weights, again, similar to what was shown in Fig.~\ref{fig:experiments}(e).

The use of such composite model's maps allows for a fast and visual inspection of anomalous diffusion data. Nevertheless, the plots presented in Fig.~\ref{fig:composite_PR}(a) shows the average of the MSE over the trajectories of the dataset. However, one of the goals of this work is to analyze the diffusion at the single trajectory level. To do so with the previous method, we look at the prediction for each trajectory with the AEs trained with composite models. Then, we find the AE that yields the lowest MSE. In Fig.~\ref{fig:composite_PR}(b) we show the percentage of trajectories whose minimal MSE is found at some $\gamma$ and $\beta$. As expected, such approach is analogous to the one of Fig.~\ref{fig:composite_PR}(a) but allows for a single trajectory based analysis. As before, we see that below the critical hormone concentration (1 nM), there is big heterogeneity in the data. As explained in Ref.~\cite{munoz2020phase}, in this cases two diffusion modes coexist, one related to the anomalous diffusion of PR through the cell membrane and the other related to phase separated PR interacting with the chromatin. Only above the 1 nM hormone concentration, the diffusion is dominated by the chromatin interaction, hence most of the trajectories have minimal MSE with AE trained with pure FBM trajectories (i.e. $\gamma=1$ and $\beta = 0$).

\section{Conclusions}

We have shown that autoencoders can learn to reconstruct trajectories arising from various anomalous diffusion models. By using the reconstruction error as a metric, we presented a method based on anomaly detection able to characterize the diffusion at the single trajectory level. Particularly, we focus on: 1) the discrimination of the theoretical diffusion model that better describes the trajectory's features; 2) the scaling factor of the mean squared displacement, defined as the anomalous exponent, usually used to evidentiate the arising of anomalous diffusion. More importantly, the training is completely unsupervised. In previous machine learning approaches, single trajectory characterization was based on either a classification problem (for diffusion models) or a regression problem (anomalous exponent). This makes the discovery of novel types of diffusion very challenging or even impossible, as the machine is restricted either to a pool of models or a range of parameters set by the training data.

Anomaly detection avoids such restrictions, as the training is solely based on the correct reconstruction of the given trajectories. Hence, any relevant information learnt by the AE is directly related to the features of the trajectories, rather than a preassigned label. This method then allows to discover novel types of diffusion, as the machine is not biased at assigning a predefined label to the trajectories. Contrarily, these new types will be heralded by an increased reconstruction error. More importantly, the method is successful at characterizing experimental trajectories, even when these are extremely short or noisy. 

This work paves the way for the study of anomalous diffusion with unsupervised machine learning. While an extended theoretical framework for anomalous diffusion exists nowadays, its connection with the phenomena arising in real physical systems is still challenging. Indeed, it is not clear to what extent the diffusion in complex scenarios, as e.g. crowded environments or the presence of non-trivial interactions, can be fully described by an idealist theoretical model. In that sense, combinations of existing theories, as the ones considered in this work, or data-driven models, may indeed give a better description of the actual physical phenomena. As shown, anomaly detection allows us to test such approaches w.r.t. to experimental data in a reliable and interpretable form.

\begin{acknowledgments}
We thank Carlo Manzo for stimulating discussions, as well as sharing experimental data; We thank Juan A. Torreno-Pina and Matthias Weiss for sharing experimental data. We also thank Pamina Winkler for useful comments on the manuscript. We acknowledge support from ERC AdG NOQIA, State Research Agency AEI (“Severo Ochoa” Center of Excellence CEX2019-000910-S, Plan National FIDEUA PID2019-106901GB-I00/10.13039 / 501100011033, FPI, QUANTERA MAQS PCI2019-111828-2 / 10.13039/501100011033), Fundació Privada Cellex, Fundació Mir-Puig, Generalitat de Catalunya (AGAUR Grant No. 2017 SGR 1341, CERCA program, QuantumCAT \ U16-011424, co-funded by ERDF Operational Program of Catalonia 2014-2020), EU Horizon 2020 FET-OPEN OPTOLogic (Grant No 899794), and the National Science Centre, Poland (Symfonia Grant No. 2016/20/W/ST4/00314), Marie Sk\l odowska-Curie grant STREDCH No 101029393, “La Caixa” Junior Leaders fellowships (ID100010434), and EU Horizon 2020 under Marie Sk\l odowska-Curie grant agreement No 847648 (LCF/BQ/PI19/11690013, LCF/BQ/PI20/11760031, LCF/BQ/PR20/11770012).
\end{acknowledgments}

\bibliographystyle{unsrt}
\bibliography{biblio}

\appendix 
\addcontentsline{toc}{chapter}{APPENDICES}
\section{Autoencoder architecture}
\label{app:ae_architecture}

\begin{table*}[ht]
    \begin{tabular}{| c | c | c | c | c |}
    \hline
     Figure & Diffusion models & Exponents & \# of trajectories  & Batch size\\ \hline
    2(b), 5(c) & ATTM & 0.1, 0.2, ..., 1 & 80.000 & 256\\ \hline
    2(c), 5(b) & CTRW & 0.1, 0.2, ..., 1 & 80.000 & 256 \\ \hline
    2(d), 3(c) & FBM & 0.1, 0.2, ..., 1 & 80.000 & 256\\ \hline
    2(e), 3(d), 5(d) & SBM & 0.1, 0.2, ..., 1 & 80.000 & 256 \\ \hline
    3(a) & FBM & 0.1, 0.2, ..., 1.9 & 160.000 & 512 \\ \hline
    3(b) & SBM & 0.1, 0.2, ..., 1.9 & 160.000 & 512 \\ \hline
    3(e) & FBM & 1, 1.1, ..., 1.9 & 80.000 & 512 \\ \hline
    3(f) & SBM & 1, 1.1, ..., 1.9 & 80.000 & 512 \\ \hline
    4(a,d) & CTRW, FBM & 0.3 & 8000 & 64 \\ \hline
    4(b,e) & CTRW, FBM & 0.6 & 8000 & 64 \\ \hline
    4(c,f) & CTRW, FBM & 0.9 & 8000 & 64 \\ \hline
    4(c,f) & CTRW, FBM & 0.9 & 8000 & 64 \\ \hline
    7 & mix (see Section~\ref{sec:composite}) & 0.8 & 10000 & 64 \\ \hline
    \end{tabular}
    \caption{Training dataset characteristics for every autoencoder used in this work. The second and third column describe the diffusion models and exponents considered for the training of the autoencoder. The number of trajectories per exponent and model is always balanced. The fourth column presents the total number of trajectories considered. In all cases, we chose a learning rate of $10^{-3}$ and a L2 regularization parameter of $10^{-5}$.}
\end{table*}

In this Appendix we present some of the techinical details related to the autoecoders used. The complete architectures as well as some examples of use can be in found in Ref.~\cite{github}. In this work, we used two different autoencoder architectures: a small autoencoder for short trajectories (20 points) and a larger one for long trajectories (200 points). In both architectures the encoder is made of 1D convolutional layers with ReLU as activation function and the decoder is made of 1D transpose convolutions without activation function. Due to the scale of the data (i.e. not a restricted range and being both positive or negative), we found that not considering any activations in the decoder yield better results.

The encoder in the small autoencoder is made of three building blocks; each building block consists of two convolutional layers with increasing number of channels, followed by a pooling operator. In the first two blocks, the pooling operation is max pooling with stride 2, which reduces the dimensions by half. In the last block, the pooling operation is global average pooling over all the channels. Batch normalization is performed after the convolutions in each building block before pooling. The last two layers of the encoder are linear layers. The last layer has 4 output neurons, meaning that the size of the latent space is 5 times smaller than the input.

The architecture of the decoder is almost identical to that of the encoder, but performs essentially the inverse operation. It has the same number of layers (2 linear layers followed by 3 building blocks). Each block has two transpose convolutional layers followed by batch normalization and max unpooling.

The large autoencoder for long trajectories (200 points) has a similar structure as the small one, the main difference being that each building block has three convolutional layers instead of two. Additionally, the output of the first layer in each block is added to the output of the third layer, following the \textit{skip connection} proposed in the ResNet architecture \cite{he2016deep}. Such feature improves dramatically the learning of deep convolutional networks with many layers and avoids the vanishing gradient problem. The extra layers and parameters allow for a better approximation of the larger trajectories. In this autoencoder the latent space has 20 neurons, 10 times smaller than the input.

The number of epochs for the training varied depending of the dataset characteristics, and was set such as to find the convergence of the MSE loss function. In general, around 200 epochs were needed. The optimization of the parameters was done via the adaptive moment estimation (Adam) method~\cite{kingma2014adam}.

\end{document}